\documentclass{iaus}
\usepackage{graphicx, epsfig}

\title[Necessary ingredients for ULXs] 
{Recipes for ULX formation: \\
 necessary ingredients and garnishments}

\author[R. Soria]   
{Roberto Soria$^{1,2}$}%

\affiliation{$^1$Harvard-Smithsonian CfA, 60 Garden st,
    Cambridge, MA 02138, USA \break email: rsoria@cfa.harvard.edu\\[\affilskip]
$^2$MSSL, University College London, 
Holmbury St Mary, Dorking, RH5 6NT, UK}

\pubyear{2006}
\volume{238}  
\pagerange{001--999}
\date{01 November 2006}
\jname{Black Holes: from Stars to Galaxies -- across the Range of Masses}
\editors{V. Karas \& G. Matt, eds.}
\begin{document}

\def\ga{\lower.5ex\hbox{$\buildrel>\over\sim$}}
\def\la{\lower.5ex\hbox{$\buildrel<\over\sim$}}

\maketitle

\begin{abstract}
I summarize the main observational features that seem 
to recur more frequently in the ULX population. 
I speculate that two of the most important physical 
requirements for ULX formation are low metal abundance, 
and clustered star formation triggered by external 
processes such as molecular cloud collisions. 
In this scenario, most ULX 
are formed from recent stellar processes, have 
BH masses $< 100 M_{\odot}$, and do not require 
merger processes in super star clusters. 
\end{abstract}

\firstsection 
\section{Common features in the ULX population}

The nature and formation mechanisms of ULXs 
remain unclear. Much of the uncertainty is due 
to the lack of direct mass estimates 
of the compact objects that power ULXs.
It is difficult to unequivocally identify  
or take spectra (let alone phase-resolved 
spectra) of their optical counterparts, 
at distances $\ga$ a few Mpc. This prevents 
the determination of ULX mass functions. 
Moreover, ULXs are an ill-defined class of systems, 
based simply on their apparent luminosity: they 
may include diverse physical objects.
Nonetheless, it is possible and useful to summarize 
common features that appear associated with a majority
of ULXs, to find out which of those phenomena are 
a clue to their physical nature. Here, we discuss 
some of them, with particular attention to the brightest 
ULXs (i.e., those with X-ray luminosities $\ga 10^{40}$ 
erg s$^{-1}$, of which only $\sim 20$ are known).

\begin{itemize}
\item {\bf Do ULXs require an accreting black hole (BH)?} 
In most cases, {\it YES}. 
Some young supernova remnants (SNRs) can be misidentified 
as ULXs (e.g., Circinus Galaxy X-2). 
So can young high-energy pulsars, with magnetic fields $\sim 10^{14}$ 
Gauss and rotational periods $\sim 10$ ms (Stella \& Perna 2004); 
they are also likely to be associated with young SNRs. 
But most ULXs show long-term fluctuations 
and flux variability inconsistent with the SNR 
or pulsar model, and lack X-ray emission lines, 
unlike typical SNR spectra.

\item {\bf Do they require BHs more massive than 
typical Galactic BHs?} That is, with masses $> 20 M_{\odot}$?
Probably {\it YES}. Their apparent luminosity is 
up to 50 times higher, and at least a few of them are inconsistent 
with strong beaming. Models explaining such a high enhancement 
entirely with super-Eddington emission or beaming
are not ruled out yet, but the simplest scenario consistent 
with the observations is to allow for higher BH masses.
If so, it means that the main difference between ULXs 
and Galactic BHs is due to the compact object, rather than 
the companion star or the gas flow.

\item {\bf Do they require intermediate-mass BHs?} 
That is, with masses $\ga 200 M_{\odot}$?
This is a controversial issue. In the absence of 
kinematic masses, various indirect methods have been suggested: 
X-ray spectral modelling, 
timing analysis, breaks in the luminosity function, 
patterns of state transitions, X-ray--radio correlations. 
Timing features such as low-frequency QPOs 
and breaks in the power-density spectrum suggest masses 
either one or two orders of magnitude higher than stellar-mass BHs, 
depending on the assumed model. The break or cutoff 
in the ULX luminosity function at $\approx 3\times 10^{40}$ 
erg s$^{-1}$ (Swartz et al.~2004; Gilfanov et al.~2004) suggests 
an upper mass limit $\approx 200 M_{\odot}$ if the Eddington limit 
is adhered to, or less if super-Eddinton emission 
is allowed. 

The presence of a soft-excess in the X-ray spectra 
of the brightest ULXs, with a characteristic temperature 
$\approx 0.15$ keV, was interpreted as evidence in favor 
of BH masses $\sim 1000 M_{\odot}$, if the emission comes 
from a standard accretion disk (Miller et al.~2004). 
However, we argued (Soria et al.~2006) that 
that argument is incorrect: when the dominant power-law 
component is also taken into account, the luminosity 
and temperature are consistent with masses $\sim 50 M_{\odot}$. 
It is also possible that the soft excess does not 
come from a disk, but from reprocessing in an ionized outflow
(Gon\c{c}alves \& Soria 2006).

In conclusion, the available observational evidence 
does {\it NOT} require intermediate-mass BHs with masses 
$\sim 1000 M_{\odot}$ (although they are not ruled out, either), 
and is still consistent with masses $\la 100 M_{\odot}$.
Stellar-evolution models predict that He cores with masses 
$64 M_{\odot} \la M \la 133 M_{\odot}$ are disrupted by 
the pair instability and do not collapse into a BH 
(Heger \& Woosley 2002; Yungelson 2006).
Therefore, there might be two subclasses of ULXs: 
one with BH masses $\la 70 M_{\odot}$ (accounting for some 
mass increase due to accretion) and one with masses   
$\ga 130 M_{\odot}$. There is no observational 
evidence of such dichotomy, and it is likely that most ULXs 
belong to the lower-mass group. However, 
a few (4 or 5) ULXs have been observed at least once with 
apparent luminosities $\sim 5$--$12 \times 10^{40}$ 
erg s$^{-1}$ 
and have been labelled ``hyperluminous X-ray sources'' 
by some authors. One may speculate 
that they belong to the higher-mass group;  
alternatively, some of them could be nuclear BHs 
of disrupted satellite galaxies (King \& Dehnen 2005).

\item {\bf Do they require young stellar environments?} 
{\it YES.} 
Most ULXs brighter than a few $10^{39}$ erg s$^{-1}$
are located in star-forming environments, rather than 
spiral bulges, halos and elliptical galaxies. 
This suggests that ULXs are scaled-up versions 
of high-mass X-ray binaries, with an OB donor star 
overflowing its Roche lobe. This enables a mass 
transfer rate $\ga 10^{-6} M_{\odot}$ yr$^{-1}$ 
for a few Myr (nuclear timescale), and suggests 
characteristic ULX ages $\sim 10^7$ yr. 
Low-mass donors could reach this level of mass transfer 
only during short-lived (thermal timescale) 
evolutionary phases, at a much later age ($> 1$ Gyr).

\item {\bf Do they require a donor star in a binary system?} 
Probably {\it YES.} Models based on Bondi accretion from molecular 
clouds (Krolik 2005) cannot be ruled out in some cases, 
but are generally disfavoured by the low absorption 
seen in almost all ULX spectra (typically $< 10^{21}$ cm$^{-2}$) 
and low extinction in the surrounding stellar population.
It is unlikely that all ULXs accreting from molecular 
clouds are at the very edge of them. 

\item {\bf Do they require starburst environments?} 
{\it NO.} 
Although 
starburst conditions are positively correlated 
with ULX formation, they do not seem to be a necessary condition. 
Some of the brightest ULXs are located in dwarf irregular 
galaxies with only localized star formation (such as 
those in Ho II and Ho IX, in the M\,81 group). 
Others are located in normal star-forming (not starburst) 
galaxies, such as NGC\,1313 and NGC\,1365. 
A few are in relatively quiescent environments 
of starburst galaxies (for example, the two brightest ULXs 
in NGC\,7714), many kpc away from the starburst region.

\item {\bf Does ULX formation require super-star-clusters (SSCs)?} 
{\it NO.} Very few ULXs are found in SSCs, 
defined as young, compact clusters with stellar masses $\ga$ a few 
$10^5 M_{\odot}$ and sizes $\la$ a few pc. Among ULXs 
with $L_{\rm X} > 10^{40}$ erg s$^{-1}$, the only 
examples are one in M\,82 and one in NGC\,7714; none  
are found in the Antennae; there may be some 
in the Cartwheel but it is too far for unequivocal 
identifications. In most cases, ULXs are near or inside 
OB associations or small open clusters, with no 
SSCs nearby. Characteristic stellar ages ($\sim 10^7$ yr) 
are too young to be consistent with the evaporation 
of a hypothetical parent SSC. Even if we assume that 
a parent cluster had time to disperse, the integrated 
mass of all the stars seen today within $\sim 100$ pc 
of a ULX does not generally add up to $10^5 M_{\odot}$; 
more typical values are $\la 10^4 M_{\odot}$.
Finally, when SSCs and ULXs are present in the same region, 
typical displacements are too large (a few hundred pc) 
to be consistent with cluster ejection.

SSCs were modelled as an ideal environment 
to form BHs as massive as $\sim 500$--$1000 M_{\odot}$ 
in the local universe, via runaway core-collapse and merger 
of O stars over a timescale $\la 3$ Myr 
(Portegies Zwart \& McMillan 2002). However, 
we have argued that there is no longer a compelling 
need to invoke intermediate-mass BHs in ULXs, 
and that the upper mass limit is likely to be 
somewhere between $50$ and $200 M_{\odot}$.
Correspondingly, if dynamical collapse and merger 
processes are still needed to form a very massive 
stellar progenitor ($> 100 M_{\odot}$), clusters 
as small as $\sim 10^4 M_{\odot}$ may do the job. 
We have also argued (Soria 2006) that collapse and merger 
processes can be more efficient at an earlier stage 
of cluster evolution, when its protostars 
are still surrounded by large, optically-thick 
envelopes, and are still accreting from neutral 
intracluster gas. Collapsing molecular clumps 
with masses $\sim 10^4 M_{\odot}$ are large 
enough to enable the formation of stars with masses 
$> 100 M_{\odot}$ via accretion and coalescence, 
and at the same time are small 
enough to disperse quickly after the most massive 
stars reach the main sequence (Kroupa \& Boily 2003), 
leaving behind an open cluster or OB association, 
in agreement with the observations. 

\item {\bf Does ULX formation require low metal abundance?}
Almost certainly {\it YES.} This is supported both by (still sketchy) 
empirical evidence and theoretical arguments. We leave 
a detailed discussion of the available metallicity 
data for ULX host galaxies to further work. It is 
of course more difficult to produce BHs at higher metallicities, 
because more mass is lost from the progenitor 
star via stellar winds. At solar-metallicity, 
all O stars --- including the Pistol star and $\eta$ Carinae, 
despite their initial masses $\approx 150$--$200 M_{\odot}$ --- are 
predicted to leave behind only a neutron star. At $Z \sim 0.1$,
they may produce BHs with masses $\sim 50 M_{\odot}$.

\item {\bf Does ULX formation require primordial abundances?} 
Probably {\it NOT.} Massive Pop-III stars
were suggested as an alternative to local SSC scenarios 
for IMBH production. There may well be Pop-III 
BH remnants with masses up to $\sim 1000 M_{\odot}$ 
floating around in galactic halos, or slowly sinking towards the
centres, but this does not explain the observed ULX correlation 
with young, star-forming environments. The Pop-III scenario 
requires that floating BH remnants
capture an OB star while they cross a star-forming environment, 
perhaps after being thrown out of their halo orbits 
during tidal interactions and collisions. In the absence 
of independent evidence for the very existence of Pop-III 
remnants, it remains an unlikely (though interesting) 
conjecture, especially if IMBHs are not needed after all 
to explain the ULX luminosity.

\item {\bf Is ULX formation directly favoured by tidal 
interactions and collisions?}
Apparently {\it YES.} Many ULXs are found in tidal dwarfs, 
or colliding galaxies, or dwarf irregular galaxies 
located in tidally interacting groups.
Spectacular examples include galaxies such as 
the Antennae, the Mice, the Cartwheel, NGC\,7714/15, 
NGC\,4485/90, and the M\,81/M\,82 group.
Other bright ULXs are associated with local collisional events: 
NGC\,4559 X-1 is in a ring of star formation 
(age $\sim 20$ Myr) probably caused by a small satellite 
galaxy splashing through the gas-rich disk;
M\,99 X-1 is apparently located where a large, fast 
H{\footnotesize {I}} cloud is impacting the outer  
disk; NGC\,1313 may have undergone 
a recent collision with a satellite, near its ULX X-2.

Are these chance associations? One simple explanation could be 
that collisions enhance star formation, and a high star-formation 
rate (SFR) leads to more X-ray binaries and a larger probability 
to form ULXs --- the normalization 
of the high-mass X-ray binary luminosity function 
being proportional to the SFR (Gilfanov et al.~2004). 
While this is probably part of the explanation, 
it cannot the whole story. In various cases, the local 
SFR in a collisional or tidal feature is small, compared 
with the SFR in the rest of the galaxy or group; and yet 
ULXs seem to be directly associated to those environments 
(NGC\,7714, M\,99 and NGC\,4559 are striking examples). 
I suggest that there can be a direct physical association 
between collisions and ULX formation, if collisions tend to produce
a {\it qualitatively different kind of star formation, that is more likely 
to lead to the formation of relatively massive BH remnants}, and hence 
to some ULXs.
\end{itemize}


\section{Outlining a plausible ULX scenario}

Taking into account the previous arguments, I speculate that 
the following line of investigation appears the most promising. 
Most ULXs contain BHs with masses $\sim 50 M_{\odot}$ and 
in any case $\la 100 M_{\odot}$, formed via direct 
core collapse from very massive stellar progenitors, 
and accrete from an OB star coeval with the BH progenitor.
The luminosity enhancement with respect 
to Galactic BHs can be explained with a factor 
of $\approx 5$--$10$ in mass, and $\approx 5$  
in super-Eddington emission, particularly outside 
the disk plane.

Progenitor stars with initial masses $\sim 150$--$200 M_{\odot}$ 
do exist (although they are very rare), and can be formed 
in clustered environments, via fast gas accretion 
and mergers of smaller protostars --- this is also 
the way ordinary O stars are thought to form.
The protocluster NGC\,2264C in the Cone nebula 
is a textbook example of a molecular clump, 
with a gas mass $\approx 1700 M_{\odot}$ that 
is undergoing dynamical collapse (infall 
of $\ga 10^{-3} M_{\odot}$ yr$^{-1}$ 
over a free-fall timescale of $\sim 10^{5}$ yr)
rather than turbulent fragmentation 
(Peretto et al.~2006).
There is no need for such clusters to be 
more massive than $\sim 10^4 M_{\odot}$.

Such global collapses occur when star formation 
is triggered by an external pressure wave.
Cloud-galaxy or galaxy-galaxy collisions 
provide ideal environment for triggered 
star formation and therefore also for 
massive stellar progenitors. Low metal abundance 
provides the second ingredient, ensuring 
that a massive BH remnant is formed.

The normalization of the high-mass X-ray binary luminosity 
function, and probably also the number of fainter ULXs 
with luminosities $\sim$ a few $10^{39}$ erg s$^{-1}$, 
is directly proportional to the SFR. However, the location 
of the upper-luminosity break, and hence the probability 
of forming ULXs with luminosities $> 10^{40}$ erg s$^{-1}$, 
depends more strongly on the two factors mentioned 
above: external triggers and low metal abundance.
A key observational test would be to map the presence 
of very massive stars in nearby galaxies, although it 
may be difficult to distinguish them from unresolved 
stellar groups. 


\end{document}